\documentclass[preprint, showpacs,preprintnumbers,amsmath,amssymb,nofootinbib]{revtex4}
\usepackage{amssymb}
\usepackage{natbib}
\usepackage{graphicx}
\usepackage{dcolumn}
\usepackage{bm}

\newcommand{\bea}{\begin{eqnarray}}
\newcommand{\eea}{\end{eqnarray}}

\begin{document}
\title{ Probing the cosmic opacity from Future Gravitational Wave Standard Sirens}
 \author{Lu Zhou$^1$, Xiangyun Fu$^1$\footnote{corresponding author:  xyfu@hnust.edu.cn}, Zhaohui Peng$^1$ and Jun Chen$^2$}
\affiliation{$^1$Institute of  Physics, Hunan University of Science and Technology, Xiangtan, Hunan 411201, China\\
$^2$School of Science, Kaili University,
Kaili, Guizhou 556011, China
}

\begin{abstract}
In this work, using the Gaussian Process, we explore the potentiality of  future gravitational wave (GW)
measurement to probe cosmic opacity through comparing its opacity-free luminosity distance (LD) with the opacity-dependent one from type
Ia supernovae (SNIa). GW data points are simulated from the third generation Einstein Telescope, and SNIa data are
taken from the Joint Light Analysis (JLA) or Pantheon compilation. The advantages of using Gaussian
Process are that one may match  SNIa data with GW data at the same redshift and use all
available data to probe cosmic opacity.  We obtain
that the error bar of the constraint on cosmic opacity can be reduced to $\sigma_{\epsilon}\sim 0.011$  and $0.006$ at $1\sigma$ confidence level (CL) for JLA and Pantheon respectively in a cosmological-independent way. Thus,  the future GW measurements  can give competitive results on the cosmic opacity test. Furthermore, we propose a method to probe the spatial
homogeneity of the cosmic transparency through comparing the reconstructed LD from the mock GW with
the reconstructed one from SNIa data in a flat $\Lambda$CDM with the Gaussian Process. The result shows that a transparent universe is favored  at $1\sigma$ CL, although the best-fit value of
cosmic opacity is  redshift-dependent.

$\mathbf{Keywords:}$  Cosmic opacity,  gravitation wave
\end{abstract}


 \maketitle

\section{Introduction}
The unexpected dimming of the type Ia supernovae (SNIa) was first attributed to the evidence of accelerating expansion of the universe~\cite{Riess1998,Perlmutter1999}. The usual explanation for this phenomena, in the frame of the General Relativity, is the existence of an exotic component with negative pressure, named as dark  energy.  Of course, there are also some other plausible mechanisms dimming the SNIa, such as  the non-conservation of the total number of photon, which may be resulted by the presence of scattering and absorption of some opacity sources~\cite{Lima2011}, or by the non-standard mechanisms, such as scalar fields coupled non-minimally with the electromagnetic Lagrangian~\cite{Hees2014,Holanda2017,Holanda20161,Aguirre1999} or oscillation of photons propagating in extragalactic magnetic fields into light axion~\cite{Csaki2002,Avgoustidis2010,Avgoustidis2009,Jaeckel2010}. Any changes in the photon flux during propagation towards the Earth will affect
the supernovae luminosity distance (LD) measures. Although the cosmic accelerating expansion has been verified by many other astronomic observations, such as large-scale structure~\cite{Tegmark2004}, baryon acoustic oscillation (BAO)~\cite{Eisenstein2005} and cosmic microwave background~\cite{Spergel2003}, cosmic opacity still needs to be taken into consideration since the cosmic acceleration rate and the cosmological parameters determined by the SNIa measurements are highly dependent on the dimming effect.  Thus,  in the era of precision cosmology, it is necessary to accurately probe the cosmic opacity with different  observational data sets and methods.

Over the past years, some tests on the  cosmic opacity have been performed by using  the  cosmic distance-duality relation (CDDR) which relates the LD $D_{\rm L}$  with the
angular diameter distance (ADD) $D_{\rm A}$ through the following identity
\begin{equation}\label{ddr}
  \frac{D_{\rm L}}{D_{\rm A}}{(1+z)}^{-2}=1.
\end{equation}
Here, $z$ is the redshift. This reciprocal relation was firstly proved by Etherington in 1933~\cite{eth1933} based on two fundamental hypotheses, namely, that light travels always along null geodesics in a
Riemannian geometry, and the number of photons  is conserved in cosmic evolution~\cite{ellis1971,ellis2007}. Many works have been performed to test the CDDR through parameterizing ${D_{\rm L}}{(1+z)}^{-2}/{D_{\rm A}}=\eta(z)$ with different astronomical observations~\cite{DeBernardis2006,Lazkoz2008,DeFilippis05,holanda2010,
holanda20103,Li2011,Boname06,Wu2015,Ellis2013,Hicken2009,Avgoustidis2010,Meng2012,Goncalves2012,Holanda20171,fuxiangyun,fuxiangyun2019}.
 Given that the Riemannian geometry is used as the mathematical tool to describe the space time of universe and photon traveling along null geodesic is more fundamental than the conservation of photon number~\cite{uzan2004,Santana2017}, any violations of  the CDDR most possibly indicate the opaque of our universe. For an opaque universe,  the photon flux received by the observers is reduced by a factor $e^{-\tau(z)}$, and the observed LD can be expressed to be~\cite{More2009},
\begin{equation}\label{ddob}
  {D_{\rm L,obs}}(z)={D_{\rm L,true}}(z)e^{\tau(z)/2},
\end{equation}
where $\tau(z)$ denotes the optical depth related to the cosmic absorption, which is related with the parametrization $\eta(z)$ for
 the CDDR through $e^{\tau(z)/2}=\eta(z)$~\cite{Lima20112}. 
  Different observational data, involving the the SNIa, the BAO,  the galaxy cluster samples, the  $H(z)$ data, the old passive galaxies and the gas mass fraction of galaxy clusters, have been used to probe the cosmic opacity~\cite{More20092, Avgoustidis2010,Avgoustidis2009, Liao2016,Holanda2014,Jesus2016,Holanda20172,Liao2011,Liao2015,Li2013}.  But, no significant opacity is obtained in these studies. Because  the presence of some opaque sources cannot be ruled out,  it remains to be necessary to employ additional tests on the cosmic opacity with various astronomical observations.

Recently, the joint detection of the gravitational wave (GW) event GW170817 with electromagnetic (EM) counterpart (GRB 170817A) from the merger of binary neutron stars (NSs)~\cite{Abbott,Abbott2,Daz,Cowperthwaite} is the first time that a cosmic event can be investigated in both EM waves and GWs,  which opens a new era of multi-messenger cosmology.
The application of GW information in cosmology was first proposed by Schuts~\cite{Schutz}, who suggested that the Hubble constant can be determined from GW observations using the fact that the waveform signals of GWs from inspiraling and merging compact binaries encode distance information. So, GW sources can be considered as standard sirens in astronomy, analogous to SNIa standard candles.   Unlike the distance estimation from  SNIa observations, one can, from the GW observations, obtain the luminosity distances directly without the need of cosmic distance ladder since stand sirens are self-calibrating. This advantage of GW measurements is that the LDs are  opacity-free. If compact binaries are NS-NS or black hole (BH)-NS binaries, the source redshift may be observed from EM counterparts that occur coincidentally with the GW events~\cite{Zhao2011,ET,Nissanke2010,Cai2017}. Thus, the LD-redshift relation can be constructed by a cosmological-model-independent way, which provides us an opportunity to make constraint on the parameters of cosmology.
Up to now, the simulated GW data have been used to measure the cosmological parameters~\cite{Zhao2011,Pozzo201217,Liao2017,Cai2016,Wei2017,Belgacem2019}, determine the total mass of neutrino~\cite{Wang2018}, investigate the anisotropy of the universe~\cite{Cai2018,Wei2018}, constrain the evolving Newton's constant G~\cite{Zhao2018}, and test the CDDR~\cite{Yang2017,fuxiangyun2019}.

More recently, Wei~\cite{JUN-JIE WEI2019}  proposed that an unbiased cosmic opacity test can be performed by combining the opacity-dependent LD from SNIa Pantheon compilation with the opacity-free LD from future observational GW data with a choice criteria ($\Delta z=z_{\rm GW}-z_{\rm SN}<0.005$) while matching the SNIa data with GW data. Then, with this choice criteria, Qi et al.~\cite{Jing-Zhao Qi2019} obtained similar investigation of cosmic opacity by combining SNIa JLA and Pantheon compilations with future GW data. Both of them found that future GW measurements will be at least competitive with current tests on cosmic opacity. However, it should be noted that one has to take care of the errors due to the mismatch between SNIa and GW data by using the choice criteria. Meanwhile, some available  data points are discarded,  since the distribution of the number density of SNIa data is different from that of the GW data in some redshift region.
Thus, if one has new method to avoid these problems,   more reliable results can be obtained, which is the motivation for our present paper.

In this work,  we employ the Gaussian Process~\cite{Seikel2012,Rasmussen2006} to reconstruct the continuous LD function from the GW standard sirens mocked with the future Einstein Telescope (ET). So, every  SNIa LD  has the corresponding  GW one with the same redshift obtained from the reconstructed GW LD function, and  then  all the available data points can be used to explore the potentiality of  future gravitational wave (GW)
measurement on probing the cosmic opacity. The advantages of the Gaussian process method are non-parametric and cosmological-model-independent, and this method  has been widely employed to reconstruct the equation of state of dark energy~\cite{Seikel2012,Cai2017} and test cosmography~\cite{Shafieloo}. It was also used to test the CDDR~\cite{Yang2017}.
In our analysis, two parameterizations: $\tau(z)=2\epsilon z$ (P1) and $\tau(z)=(1+z)^{2\epsilon}-1$ (P2) are adopted to describe the cosmic optical depth. In addition,  we obtain the continuous function of cosmic optical depth $\tau{(z)}$ through comparing the reconstructed LD from GW with the reconstructed one from the JLA SNIa data.   Our results show that, compared with the current analyses, measurements of  future GW events are very powerful  to  probe the cosmic opacity.

\section{SNIa samples and simulated GW data}
In order to explore the potentialities of future GW measurements on testing the cosmic opacity, we compare the observed LD obtained from the JLA~\cite{SDSS2014} or Pantheon~\cite{Scolnic2018} SNIa compilation with the opacity-free one from the mocked measurements of future GW events.
\subsection{JLA and Pantheon SNIa data}
JLA SNIa compilation contains
a set of 740 spectroscopically confirmed data points compiled by the Sloan Digital Sky Survey(SDSS)-II supernova survey~\cite{SDSS20142}, the Supernova Legacy Survey (SNLS) survey~\cite{SNLS2010} and a few from Hubble Space Telescope SNIa measurements~\cite{Riess2007} in the redshift range $0.01<z<1.3$. With the light-curve parameters, the distance modulus of a SNIa has the form
\begin{equation}\label{dismodu}
  \mu=m_{\rm B}-(M_{\rm B}-\alpha\times x+\beta\times c),
\end{equation}
where $m_{\rm B}$ is the rest-frame peak magnitude in the B
band, $x$ is the time stretching determined by the shape of the SNIa light curve and $c$ is the supernova color measurement at maximum brightness. These parameters are all derived from the observed light-curve, and thus are independent of any cosmological model. $\alpha$ and $\beta$ characterize the shape and color corrections of the light-curve, and  $M_{\rm B}$ is absolute magnitude of the SNIa. These three nuisance parameters are assumed to be constants for all the SNIa. 
 It has been found that $(\alpha, \beta, {\rm M}_{\rm B})=(0.141\pm 0.006, 3.101\pm 0.075,19.05\pm 0.02)$ at $1\sigma$ CL~\cite{SDSS2014} by fitting the JLA data with the flat $\Lambda$CDM model, and $(\alpha, \beta, {\rm M}_{\rm B})=(0.141\pm 0.006, 3.102\pm 0.075,19.10\pm 0.03)$ by fitting the Plank+WP+JLA+BAO data with a model which has a  constant equation of
state parameter (wCDM). In this paper, we firstly adopt these fitted values of  $(\alpha, \beta, {\rm M}_{\rm B})$ from a flat $\Lambda$CDM model to obtain the LD ($D_{\rm L}$) of SNIa through the relation $\mu(z)=5\log_{10}(D_{\rm L}(z))+25$ with the uncertainty  of  $D_{\rm L}$ being
 \begin{equation}\label{sigdis}
  \sigma_{D_{\rm L}}={\log_{10}\over 5}D_{\rm L}\sigma_\mu.
\end{equation}
The obtained LD of SNIa is shown in the left panel of Fig.~(\ref{mock1}). It is obvious that the method used to probe cosmic opacity is cosmological-model-dependent (Named Method A). In addition, in order to obtain the LD of SNIa with a model-independent way, we directly take the observational quantities ($m_{\rm B}, x, c $) and their errors, and treat the nuisance parameters $(\alpha, \beta, {\rm M}_{\rm B})$ as additional parameters. Then we  assume them distributing uniformly  over appropriate prior ranges $0.120<\alpha<0.160, 2.850<\beta<3.330$ and $18.90<{\rm M}_{\rm B}<19.20$ which is a little larger than the ranges obtained from the $\Lambda$CDM model and wCDM model at $3\sigma$ CL (named Method B).

Furthermore, we also take into consideration  the  Pantheon SNIa compilation released by  the Pan-STARRS1 Medium Deep Survey, which  consists of 1048   data points up to redshift $z\sim 2.26$. The LD distances of the Pantheon compilation are calibrated from the SALT2 light-curve fitter through applying  the BEAMS with Bias Corrections  method to determine the nuisance parameters and taking account of the distance bias corrections~\cite{Scolnic2018}. Thus,   we obtain the distance modulus from $\mu=m_{\rm B}-M_{\rm B}$ by considering the nuisance parameters ${\rm M}_{\rm B}$ to  distribute uniformly over prior range $18.90<{\rm M}_{\rm B}<19.20$.

\subsection{simulated GW data}

We simulate the GW data based on the ET. The ET is the third generation of ground based GW detector which will be able to detect GW signals to be ten times more sensitive in amplitude than the advanced ground-based detectors.  It can cover a wide range of frequency $1\sim 10^4$ Hz up to redshift $z\sim2$ for the NS-NS  and $z
\sim5$ for the BH-NS mergers systems.  The strategy implemented with the future GW detectors has been discussed in Refs.~\cite{Regimbau2012,Regimbau2014,Regimbau2015, Regimbau20152,Regimbau2016,Regimbau2017}. The important question in data analysis is whether overlapping signals can
be discriminated by the future detectors in a  GW signal-rich environment.  It has been shown that algorithms currently used to analyze LIGO and Virgo data are already powerful enough to detect the sources expected in the future GW detectors~\cite{Regimbau2012,Regimbau2014,Regimbau2016,Regimbau2017}.  In the following, we will summarize  the process of simulating the GW samples.

The interferometers of ET GW detector is sensitive to the relative difference between two distances, named strain $h(t)$.    The strain ($h(t)$), in the transverse-traceless gauge, is written as
\begin{align}
h(t)=F_+(\theta,\phi,\psi)h_+ + F_\times(\theta,\phi,\psi)h_\times \,,
\label{equa:ht}
\end{align}
where $F_{+,\times}$ is the beam pattern functions, $\psi$ denotes the polarization angle, and $\theta,\phi$ are the angles which present the location of the source relative to the detector. Following the analysis of Refs.~\cite{Zhao2011,Regimbau2012}, the antenna beam pattern functions of the ET can be written as
\begin{align}
F_+^{(1)}(\theta,\phi,\psi)={\sqrt{3}\over {2}}[{1\over {2}}(1+\cos^2(\theta))\cos (2\phi)\cos(2\psi)-\cos(\theta)\sin(2\phi)\sin(2\psi)]\, ,  \nonumber\\
F_\times^{(1)}(\theta,\phi,\psi)={\sqrt{3}\over {2}}[{1\over {2}}(1+\cos^2(\theta))\cos (2\phi)\cos(2\psi)+\cos(\theta)\sin(2\phi)\sin(2\psi)]\, .
\label{equa:FF}
\end{align}
Considering the fact that the three interferometers have $60^\circ$ with each other, fitted in an equilateral triangle, the other two antenna pattern functions can also be derived from above equation, such as $F_{+,\times}^{(2)}(\theta,\phi,\psi)=F_{+,\times}^{(1)}(\theta,\phi+2\pi/3,\psi)$ and $F_{+,\times}^{(3)}(\theta,\phi,\psi)=F_{+,\times}^{(1)}(\theta,\phi+4\pi/3,\psi)$.

Following Refs.~\cite{Zhao2011,Cai2017,Wei2018,Yang2017,Regimbau2012}, the GW signals are from the merger of binary systems with component masses $m_1$ and $m_2$. $M=m_1+m_2$ is the total mass. The chirp mass is defined as $M_{\rm c}=M \eta^{3/5}$, where $\eta=m_1m_2/M^2$ represents the symmetric mass ratio, while the observed chirp mass can be written as $M_{\rm c,\,obs}=(1+z)M_{\rm c,\, phys}$.  Assuming that the change of orbital frequency over a single period is negligible, for the waveform of GW, the stationary phase approximation is applied to compute the Fourier transform $\mathcal{H}(f)$ of the time domain waveform $h(t)$
\begin{align}
\mathcal{H}(f)=\mathcal{A}f^{-7/6}\exp[i(2\pi ft_0-\pi/4+2\psi(f/2)-\varphi_{(2.0)})],
\label{equa:hf}
\end{align}
where the Fourier amplitude $\mathcal{A}$ is given by
\begin{align}
\mathcal{A}=&~~\frac{1}{D_{\rm L}}\sqrt{F_+^2(1+\cos^2(\iota))^2+4F_\times^2\cos^2(\iota)}\nonumber\\
            &~~\times \sqrt{5\pi/96}\pi^{-7/6}\mathcal{M}_{\rm c}^{5/6}\,,
\label{equa:A}
\end{align}
 ${D_{\rm L}}$ is the LD which can be derived from GW signals. 
   $t_0$ presents the epoch of the merger,  $\iota$ is the angle of inclination   and $\varphi_{(2.0)}$  is the phase parameter.   For the simulations, we adopt the flat $\Lambda$CDM as the fiducial cosmological  model with the model parameters being
\begin{align}
h_0=0.70,~~~\Omega_{\rm m}=0.295,
\label{equa:A1}
\end{align}
which are from the constraints of the SNIa JLA compilation~\cite{SDSS2014}. Here,
 $\Omega_{\rm m}$  denotes the present
dark matter density parameter.

Given the waveforms of the GWs, the signal-to-noise ratio (SNR) for the network of three independent ET interferometers has the form
\begin{align}
\rho=\sqrt{\sum_{i=1}^{3}\langle\mathcal{H}^{(i)},\mathcal{H}^{(i)}\rangle}\,,
\label{equa:SNR}
\end{align}
where the inner product is defined as the following equation
\begin{align}
\langle a,b\rangle=4\int_{f_{\rm lower}}^{f_{\rm upper}}{\widetilde{a}(f)\widetilde{b}^{\ast}(f)+\widetilde{a}^{\ast}(f)\widetilde{b}(f)\over{2}}{df\over{S_h(f)}}\,,
\label{equa:SNR}
\end{align}
and here ${S_h(f)}$ is the one-side noise power spectral density characterizing the performance of a GW detector. $f_{\rm lower}$ represents the  lower cutoff frequency which  is fixed at 1 Hz, and $f_{\rm upper}$ does the upper cutoff one decided by the last stable orbit (LSO),   and $f_{\rm upper}=2f_{\rm LSO} $, where $f_{\rm LSO}=1/(6^{3/2}2\pi M_{\rm obs})$ is the orbit frequency at the last stable orbit. In this paper, the catalogues of NS-NS systems and NS-BH systems are simulated with the masses of NS and BH sampled by uniform distribution in the intervals $[1,2] M_\odot$ and  $[3,10] M_\odot$ respectfully. The ratio between NS-NS and BH-NS binaries is taken to be 0.03, and the signal is identified as a GW events, only if the ET interferometers have a network SNR of $\rho >8.0$ , as being presented in the Advanced LIGO-Virgo network~\cite{Abadie2010a,Zhao2011}.

The instrumental uncertainty of the LD  from GW can be obtained by using the Fisher information matrix.   It is  supposed that the LD $D_{\rm L}$ is independent of the remaining GW parameters (the inclination angle $\iota=0$), $\mathcal{H}\propto D_{\rm L}^{-1}$, and the corresponding instrumental uncertainty  has the form
\begin{align}
\sigma_{D_{\rm L}}^{\rm inst}\simeq{2D_{\rm L}\over {\rho}}\,.
\label{equa:instr}
\end{align}
The lensing uncertainty from the weak lensing can be modeled as $\sigma_{D_{\rm L}}^{\rm lens}=0.05z D_{\rm L}$~\cite{Cai2017,Zhao2011,Belgacem2019}. Therefore, the uncertainty budget of DL for GW measurements can be obtained from the following expression
\begin{align}
\sigma_{D_{\rm L}}&=\sqrt{(\sigma_{D_{\rm L}}^{\rm inst})^2+(\sigma_{D_{\rm L}}^{\rm lens})^2}\nonumber \\
&= \sqrt{\bigg({2D_{\rm L}\over {\rho}}\bigg)^2+(0.05zD_{\rm L})^2}\,.
\label{equa:errorall}
\end{align}

The redshift distribution of GW source observed on Earth can be expressed as~\cite{Sathyaprakash2010}
\begin{align}
P(z)\propto{4\pi d^2_{\rm C}(z)R(z)\over{H(z)(1+z)}}\,,
\label{equa:distri}
\end{align}
where $H(z)$ represents the Hubble parameter from the fiducial $\Lambda$CDM, $d_{\rm C}$ is co-moving distance, and $R(z)$ is the redshift-dependent merger rate of binary systems, which is taken as~\cite{Cai2017,Zhao2011}
\begin{equation}
R(z)=\begin{cases}
1+2z, & (z\leq 1) \\
\frac{3}{4}(5-z), & (1<z<5) \\
0, & (z\geq 5).
\end{cases}
\label{equa:rz}
\end{equation}

 A key question is that how many GW events with EM counterparts can be detected per year for the future ET.
  The expected rates of  NS-NS and binary BH-NS binary detections per year for the ET are about the order $10^3-10^7$~\cite{ET}. As predicted by Cai and Yang~\cite{Cai2017,Zhao2011,Yang2017} with assumption of the middle detection rate around $10^5$, about $10^2$ GW measurements with EM counterparts (duty cycle of 100\% and the efficiency $\sim 10^{-3}$ of the total number of binary coalescence) will be observed per year.  More recently, with a more realistic scenario for the detection of EM counterpart, Belgacem et al. also estimated that the number of GW BNS events with EM counterparts will be about $39\sim 51$ per year~\cite{Belgacem2019}, if the GW events are detected by third generation networks  data (assuming a 80\% duty cycle and the network SNR threshold $\rho_{\rm thresh}=12$  for the GW detector)~\cite{Regimbau2015,Regimbau2017} and the corresponding GW-GRB coincidences are obtained by assuming a GRB detector with the characteristics of the X-Gamma ray Imaging Spectrometer (XGIS).

 In this paper, following the process from Ref.~\cite{Zhao2011,Yang2017,Cai2017},
 we  simulate 500 data points in the redshift range $0<z<2.3$, 
 and show the results in Fig.~(\ref{mock1}). One can easily find that the number density of JLA SNIa data is much larger than that of the GW in the redshift range $z<0.3$, and vice versa in the redshift range $z>1.0$. So, if matching the GW data with the SNIa data by using the choice criteria ($\Delta z=z_{\rm GW}-z_{\rm SN}<0.005$),  many SNIa data points should be discarded. In order to employ all data to probe the cosmic opacity, we reconstruct the continuous LD function from the mock GW data with the Gaussian Process.

  \section{Gaussian Process}

 Gaussian Process is a  non-parametric smoothing technique used to reconstruct a continuous function from the observed data.  At each redshift point $z$, the reconstructed function is also a Gaussian distribution with a mean value and Gaussian error bands.  The outcome of observational data points at any two redshifts $z_i$ and $z_j$ are correlated through a covariance function $\kappa(z_i,z_j)$ due to their nearness to each other. This covariance function depends on  a set of hyperparameters, and there is a wide range of possible candidates for it. As the LD function versus redshift $z$ is expected to be infinitely differentiable, we use the squared exponential covariance function:
\begin{equation}
\label{cov}
\kappa(z_i,z_j)=\sigma^2_f \exp\bigg[-{(z_i-z_j)^2\over{2 l^2}}\bigg],
\end{equation}
where $\sigma_f$ and $l$ are two hyperparameters.  $\sigma_f$ gives the output variance and fixes the overall amplitude of the correction in the $y$-direction, and $l$ denotes the measure of the coherence length scale of the correlation in the $x$-direction. This square exponential kernel is the simplest default kernel for Gaussian Process. One can calculate the values of the hyperparameters by maximizing the corresponding marginal log-likelihood probability function of the distribution. The detailed description and analysis of the Gaussian Process can be found in refs.~\cite{Seikel2012,Seikel20121}.  With the Gaussian method, we obtain the continuous function of the LD from the simulated  GW data, which is shown in the right panel of Fig. (1).

\section{Method}
The most straightforward method to probe the cosmic opacity is to compare the observed LDs with the opacity-free ones at the same redshifts. The relation between the observed LD from SNIa and the true one from GW can be presented by ${D_{\rm L,SN}}(z)={D_{\rm L,GW}}(z)e^{\tau(z)/2}$. Thus, the distance modulus of SNIa can be expressed as
\begin{equation}\label{opacity}
  {\mu_{\rm SN}}(z)=5\log_{10}{D_{\rm L,GW}}(z)+25+2.5(\log_{10}e){\tau(z)}.
\end{equation}
The universe is transparent while   $\tau=0$. All deviations from a transparent universe, which occur possibly at some redshifts, will be encoded in the function $\tau{(z)}$. In our analysis, we consider
two different  parameterizations for the $\tau{(z)}$: $\tau(z)=2\epsilon z$  and $\tau(z)=(1+z)^{2\epsilon}-1$.

By comparing the distance modulus given in Eq.~(\ref{opacity}) with the observed ones from JLA or Pantheon, one can obtain the probability density of $\epsilon$ and $\xi$  through $P(\epsilon,\xi)=A\,{\rm exp}[-\chi^2(\xi, \epsilon)/2]$, where $\xi$ denotes the nuisance  parameters $\alpha$, $\beta$, ${\rm M}_{\rm B}$ for JLA (${\rm M}_{\rm B}$ for Pantheon compilation), and $A$ is a normalized coefficient, which makes $\int\int {P(\xi, \epsilon)}d\xi d\epsilon=1$. The $\chi^2(\xi, \epsilon)$ has the form
\begin{equation}
\label{chi3}
\chi^{2}(\xi, \epsilon) = \sum\frac{{\left[\tau(z,\epsilon)-
\tau_{\rm obs}(\xi) \right] }^{2}}{\sigma^2_{\tau_{\rm obs}}}\,.
\end{equation}
Here
\begin{equation}
\label{SGL}
\sigma^2_{\tau_{\rm obs}}=2\left[\left({\sigma_{D_{\rm SN}(z)}\over{D_{\rm SN}(z)}}\right)^2+\left(\sigma_{D_{\rm GW}(z)}
\over{D_{\rm GW}(z)}\right)^2\right]\, ,
\end{equation}
which is the error of   $\tau_{\rm obs}$.
When obtaining the probability distribution function of $\epsilon$,  we treat  $(\alpha, \beta, {\rm M}_{\rm B})$ as the nuisance parameters,  which are marginalized with the uniform distributions $0.120<\alpha<0.160, 2.850<\beta<3.330$ and $18.90<{\rm M}_{\rm B}<19.20$ shown as in Section II.

\begin{figure}[htbp]
\includegraphics[width=8cm]{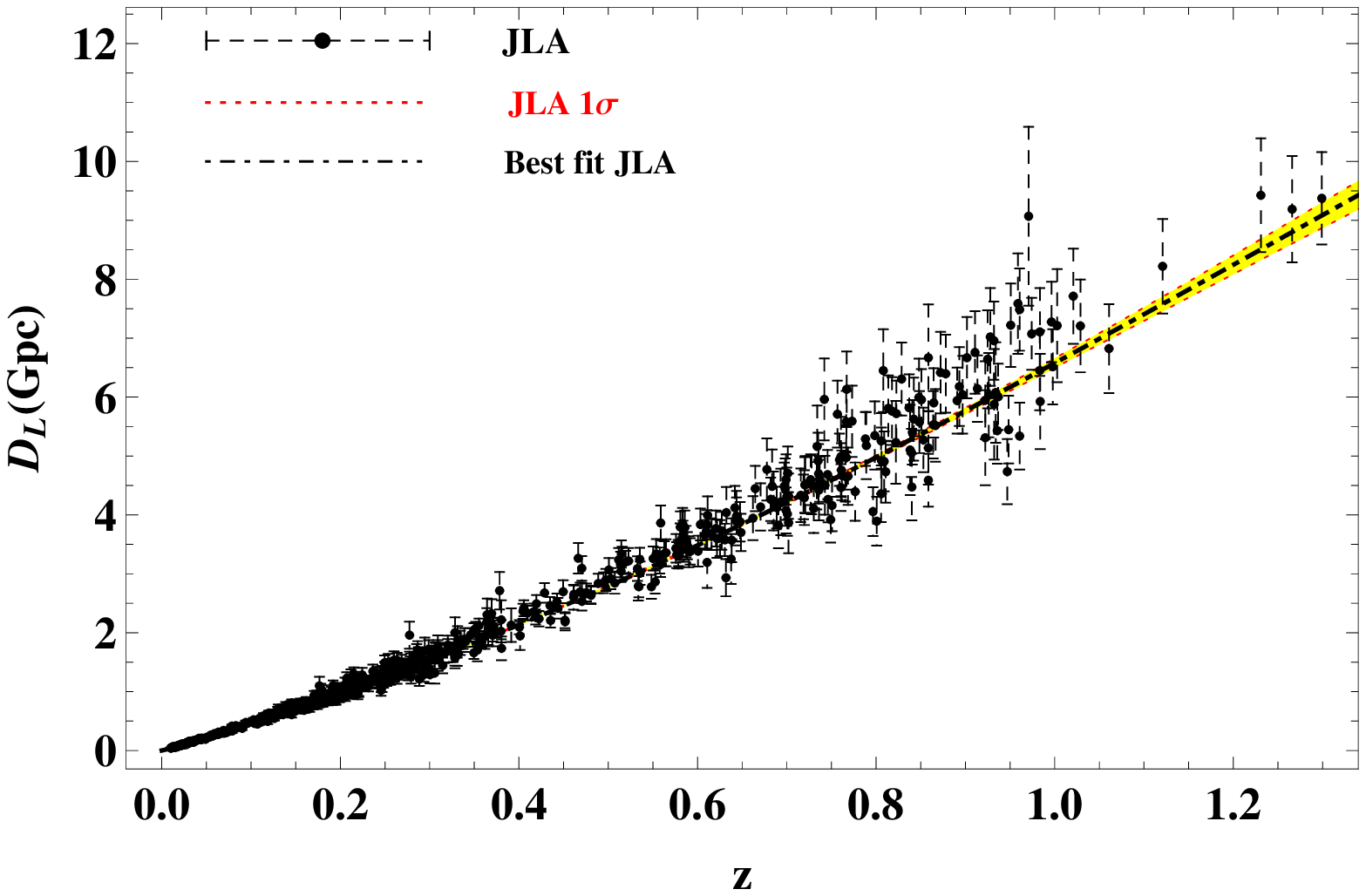}
\includegraphics[width=8cm]{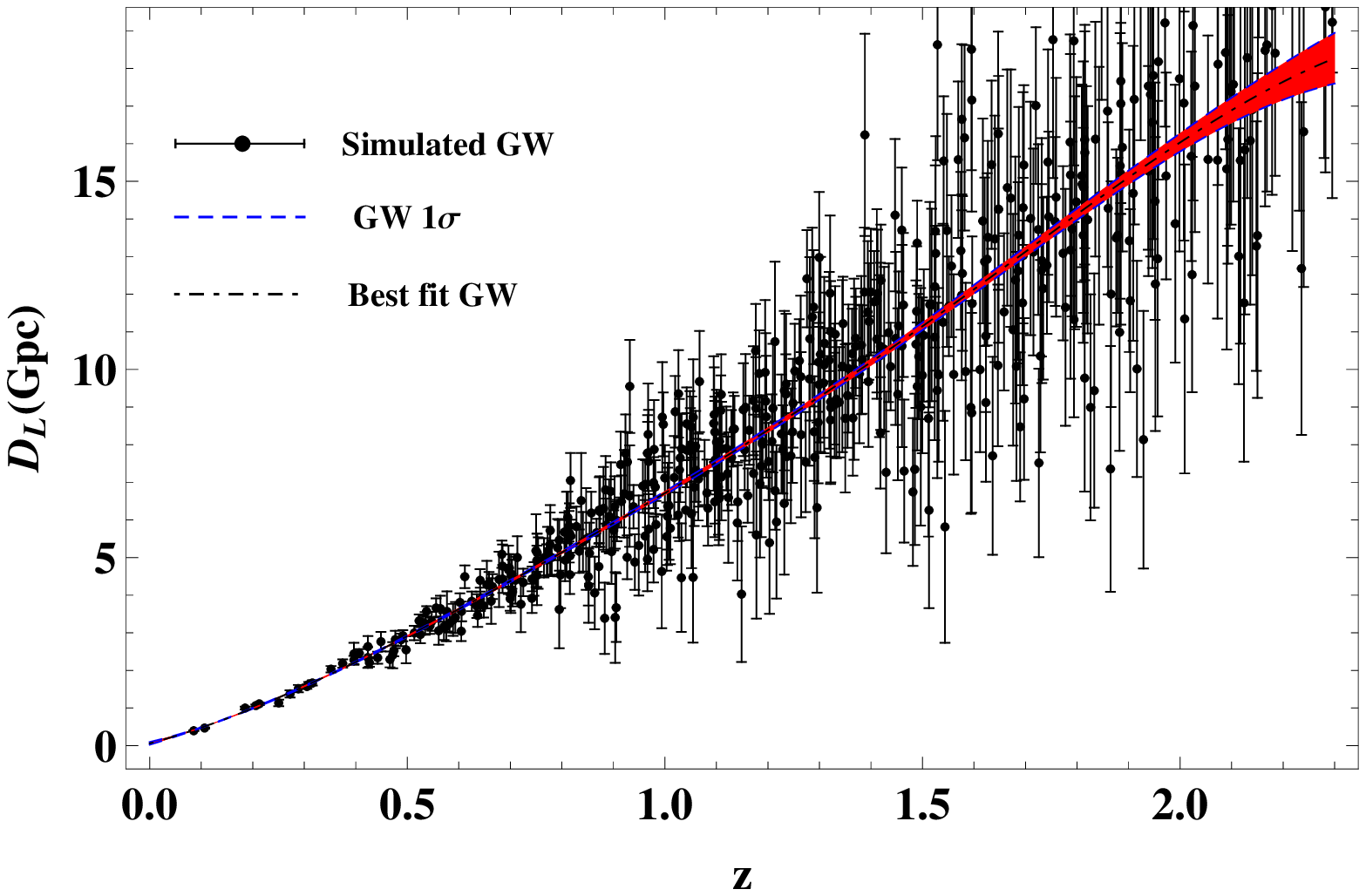}
\caption{\label{mock1} The sample catalogues of the observed JLA SNIa from $\Lambda$CDM model (left panel) and   500 mock GW events of redshifts (right panel). The corresponding smoothed LD with Gaussian Process method is also listed.}
\end{figure}

\section{Results}

\begin{table}[htp]
\begin{tabular}{|c|c|c|c|}
\hline
\scriptsize{Data }  & \   $\epsilon\,\,\,( P_1)$\ \ &$\epsilon \,\,\,( P_2)$ \\
\hline
\scriptsize{ET $\times$ JLA  \tiny{(A$^\ast$)}} &  \scriptsize{${0.004{\pm{0.008}}}$} & \scriptsize{${0.002{\pm{0.010}}}$} \\
\scriptsize{ET $\times$ JLA \tiny{(B$^\ast$)}} & \scriptsize{${0.009{\pm{0.011}}}$}&\scriptsize{ ${0.008{\pm^{0.014}_{0.012}}}$}   \\
\scriptsize{ET $\times$ Pantheon \tiny{(B$^\ast$)}} & \scriptsize{${0.005{\pm{0.006}}}$}&\scriptsize{ ${0.006{\pm{0.008}}}$}   \\
\hline
\scriptsize{H(z) $ \times$ Union \tiny{(A)} \cite{Avgoustidis2009}} & \scriptsize{${-0.01\pm^{0.08}_{0.09}}$}& $\Box$\\
\hline
\scriptsize{ H(z)$\times$ Union \tiny{(A)} \cite{Avgoustidis2010}} & \scriptsize{${-0.04\pm^{0.08}_{0.07}}$}& $\Box$\\
\hline
\scriptsize{H(z)$\times$ Union2.1\tiny{(A)} \cite{Holanda2014}} & \scriptsize{${0.06\pm^{0.18}_{0.18}}$}& $\Box$\\
\hline
\scriptsize{Clusters  $\times$ Union2.1 \tiny{(B)} \cite{Li2013}} & \scriptsize{${0.009\pm^{ 0.059}_{0.055}}$}& \scriptsize{${0.014\pm^{ 0.071}_{0.069}}$}   \\
\hline
\scriptsize{H(z)  $\times$ Union2.1\tiny{(B)} \cite{Liao2011}} & \scriptsize{${-0.01\pm 0.10}$}& \scriptsize{${-0.01\pm 0.12}$}   \\
\hline
\scriptsize{ H(z)$\times$ JLA \tiny{(B)} \cite{Liao2015}} & \scriptsize{${0.07\pm^{0.107}_{0.121}}$}& $\Box$\\
\hline
\scriptsize{ages of old objects $\times$ Union2.1 \tiny{(B)} \cite{Jesus2016}} & \scriptsize{${0.016\pm^{0.078}_{0.075}}$}& $\Box$\\
\hline
\scriptsize{ET $\times$ JLA \tiny{(B)} \cite{Jing-Zhao Qi2019}} & \scriptsize{${0.002\pm 0.035}$}&\scriptsize{ ${-0.006\pm 0.053}$}   \\
\scriptsize{ET $\times$ Panthoen \tiny{(B)} \cite{Jing-Zhao Qi2019}} & \scriptsize{${0.009\pm 0.016}$}&\scriptsize{ ${0.015\pm 0.025}$}   \\
\hline
\scriptsize{ET $\times$ Panthoen \tiny{(B)} \cite{JUN-JIE WEI2019}} & \scriptsize{${0.006\pm 0.029}$}& $\Box$   \\
\hline
\end{tabular}
\caption{The summary of maximum likelihood estimation results of $\epsilon$ for two parameterizations respectively obtained from different observations. The $\epsilon$ is represented by the best fit value at 1 $\sigma$ CL for each data set. The superscripts $\ast$ represent the results  obtained from Gaussian Process method in this work.}
\label{likelihood1}
\end{table}

\begin{figure}[htbp]
\includegraphics[width=8.8cm]{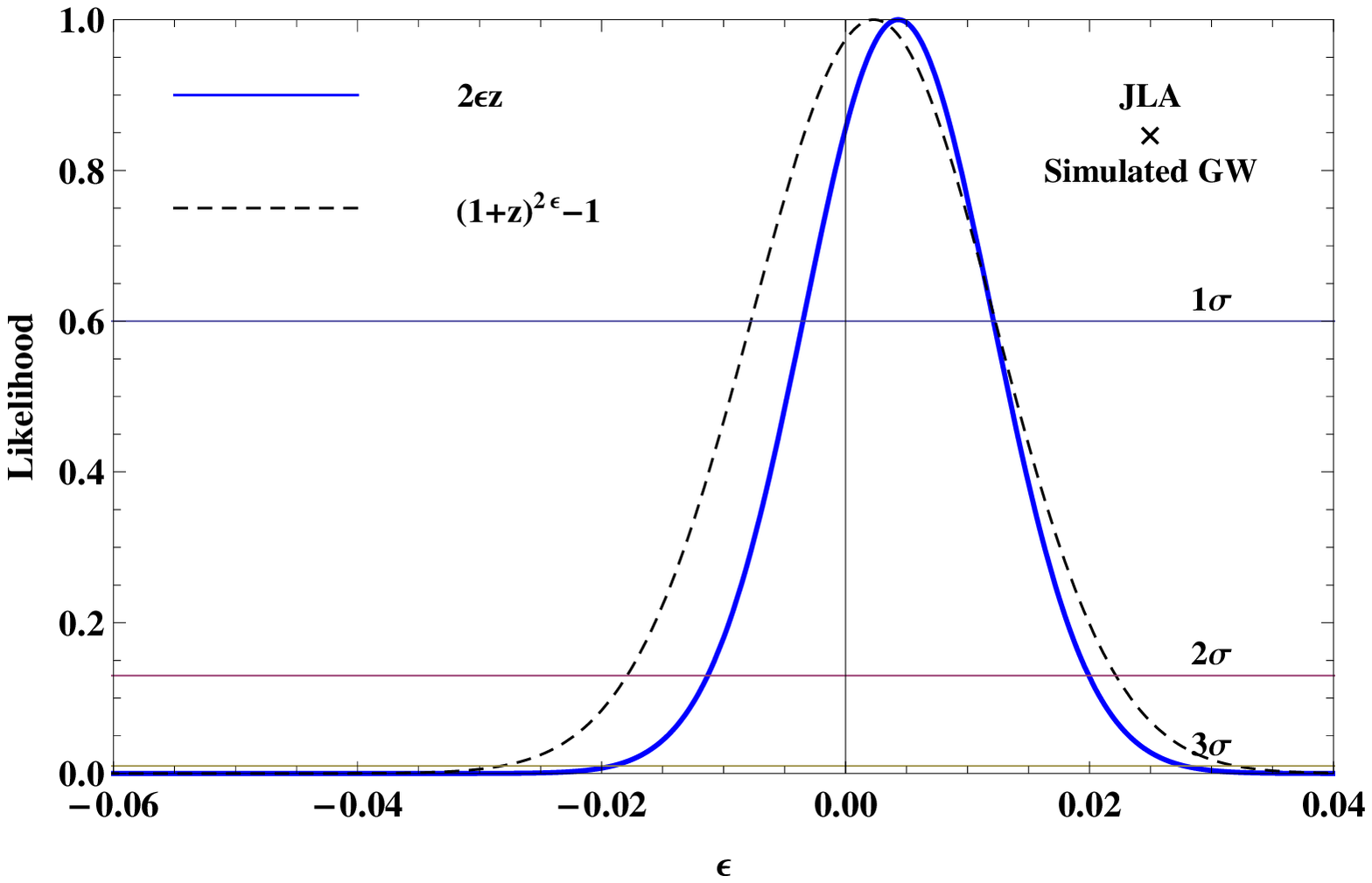}\\
\includegraphics[width=8.8cm]{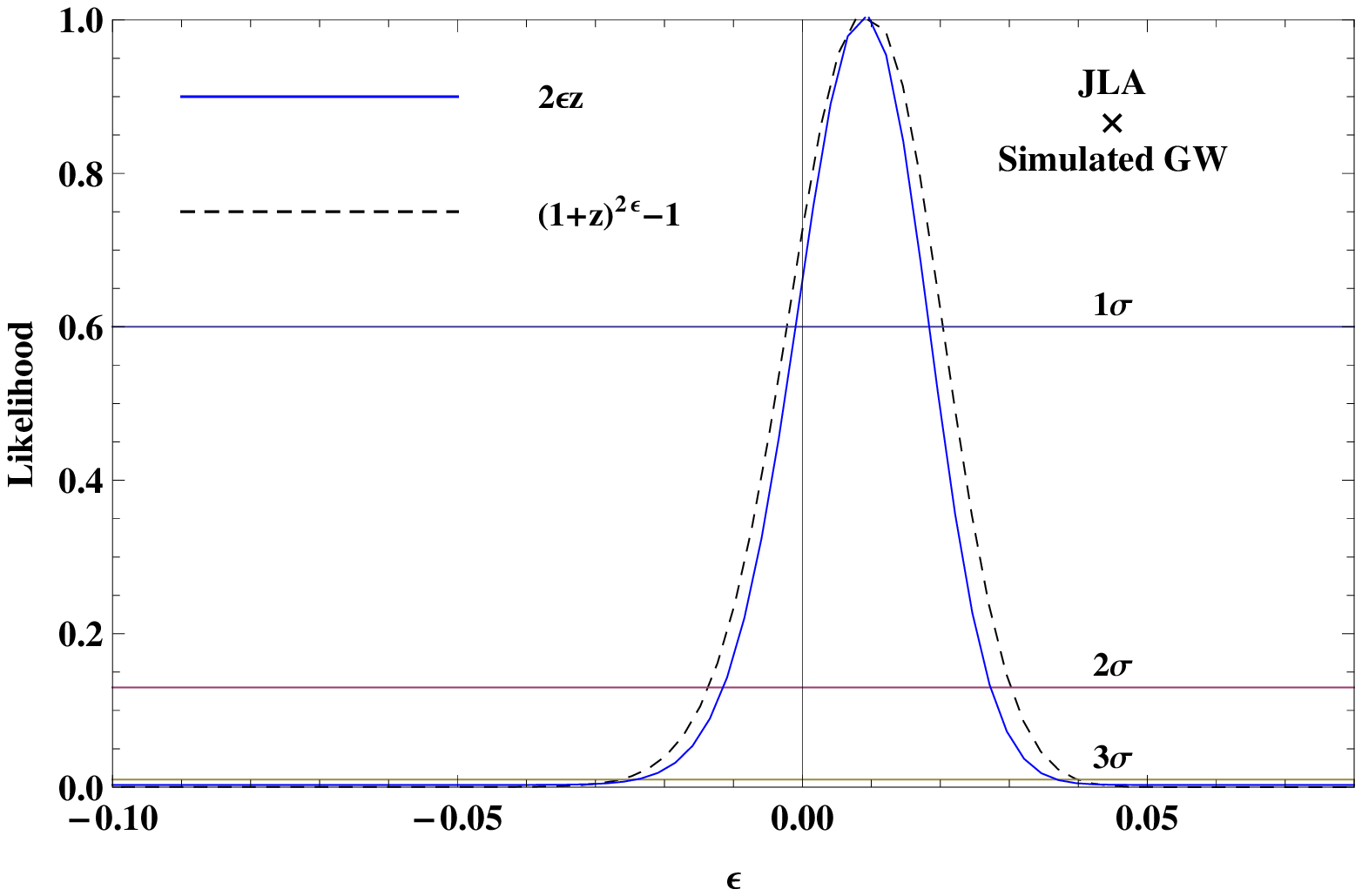}\\
\includegraphics[width=8.8cm]{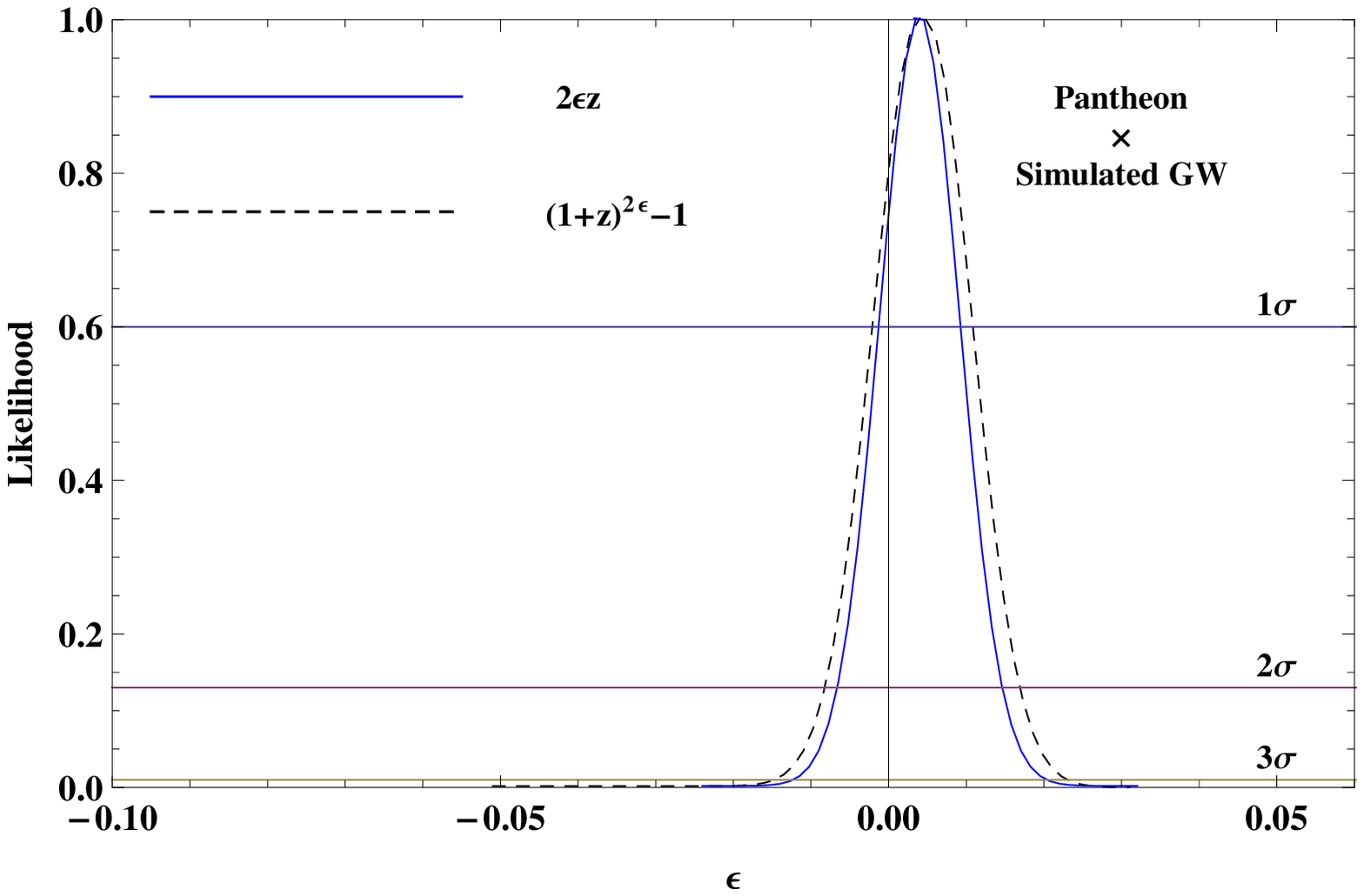}

\caption{\label{Figlikec} The likelihood  distribution functions obtained from JLA with the cosmological-model-dependent method (top), from JLA compilation with the model-independent method (middle), and from Pantheon compilation with the model-independent way (bottom) respectively. }
\end{figure}

Our results  are shown in Fig.(\ref{Figlikec}) and
Tab.(\ref{likelihood1}).
One can see that the results from the cosmological-dependent method (A) and the cosmological-independent one (B) are consistent with a transparent universe at $1\sigma$ CL. The error bar from method B is about $20\%$ larger than that from method A, as  there are more free variables  in method B. Furthermore,  the results are almost independent on the parameterizations for $\tau(z)$, although P1 may offer a stricter constraint on cosmic opacity than P2.

In Table~\ref{likelihood1},  we also give a  comparison between our forecast results and the previous tests. 
It is easy to see that all data support a transparent universe. However, our results are the tightest
constraints on the cosmic opacity.  The error bars  of our results are  at least  $80\%$
smaller than that from  the combination of the Union (Union2.1) SNIa data and the
Hubble parameters~\cite{Avgoustidis2009,Avgoustidis2010,Holanda2014,Liao2011,Liao2015},  the  Union2.1 SNIa
compilation and the galaxy cluster data~\cite{Li2013}, and the Union2.1 SNIa data and the ages of old objects~\cite{Jesus2016}.
In addition, the error bars are nearly 60\% smaller than the ones of Refs.~\cite{JUN-JIE WEI2019,Jing-Zhao Qi2019} from  the simulated ET GW data and the SNIa data under the choice criteria $z_{\rm GW}-z_{\rm SN}<0.005$.
 It indicates that both the errors from the mismatch between  two kinds of observational data compilations and the available  data points discarded by using the choice criteria  cannot be neglected.

\begin{figure}[htbp]
\includegraphics[width=10cm]{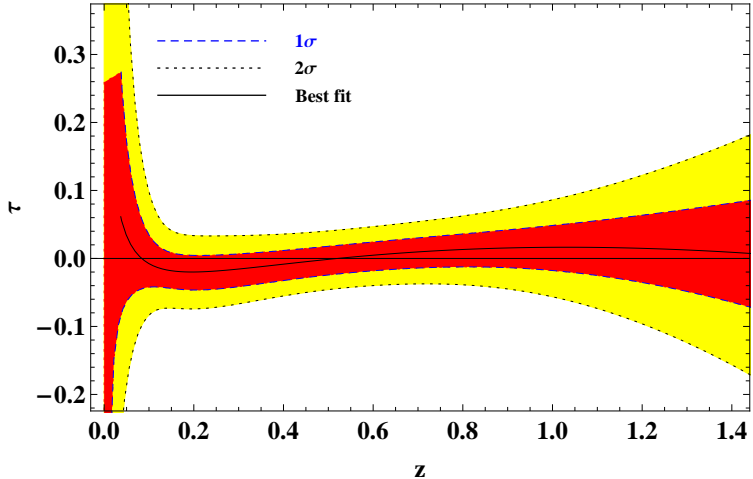}
\caption{\label{figtao} The distribution of cosmic optical depth paramter $\tau(z)$ obtained with a cosmological-dependent way.}
\end{figure}

In addition, we  reconstruct the LD ($D_{\rm L}(z)$) as a smooth function of redshift $z$ from the JLA SNIa data points with the Gaussian Process, while the distance modulus are determined with the fitted nuisances parameters (used in section II) from a flat $\Lambda$CDM model~\cite{SDSS2014}, and the reconstructed results are shown in the left panel of  Fig.~(\ref{mock1}).  Then, following the Eq.~(\ref{opacity}), the cosmic optical depth can be presented as
\begin{equation}\label{opcity2}
{\tau(z)}= [{\mu_{\rm SN}}(z)-5\log_{10}{D_{\rm L,GW}}(z)-25]/(2.5\log_{10}e).
\end{equation}
With the reconstructed results from the mock GW and SNIa data, we can obtain the continuous function of cosmic optical depth $\tau(z)$ with redshift $z$, and the results are shown in Fig.~(\ref{figtao}). The divergence of $\tau(z)$ in the redshift range $z<0.1$ may be resulted from the absence of the mock GW data. From this figure, one can find that the best-fit value of cosmic optical depth $\tau(z)$ monotonically increases in the redshift range $0.15<z<0.95$, and decreases in the range $0.95<z<1.40$. Therefore, it suggests that the cosmic opacity might be redshift-dependent, however no deviation from a  transparent universe is found  at $1\sigma$ CL. Our result is similar to the one from Ref.~\cite{Chen2012}, in which the best-fit cosmic opacity oscillates between zero and some nonzero values as the redshift varies.

\section{conclusion}
In General relativity, gravitational wave (GW) propagates freely in a perfect fluid without any absorption and dissipation,
thus the luminosity distance (LD)
measurement of GW provides us with an opportunity to probe cosmic opacity. More recently,  cosmic opacity is successively tested  by combining the opacity-dependent LD from SNIa data with the opacity-free LD from future observational GW data with a choice criteria ($\Delta z=z_{\rm GW}-z_{\rm SN}<0.005$) in Refs.~\cite{JUN-JIE WEI2019,Jing-Zhao Qi2019}, and the results show that future GW measurements will be at least competitive with current tests on cosmic opacity. However, it should be noted that some bias may be resulted by the mismatch of redshift between SNIa and GW data, and some available  data points should be discarded, while the choice criteria is used to match the GW data with the SNIa data. Thus, in order to obtain more reliable results, in this work,  we employ Gaussian Process to probe the cosmic opacity through comparing  LD from GW with the one from type Ia supernovae (SNIa). 500 GW data points are simulated, and the SNIa data are taken from the Joint Light Analysis (JLA) and Pantheon  compilation. The purpose of using Gaussian Process is to reconstruct the continuous function of LD from mock GW measurements. Then, every  SNIa data point  has the corresponding  GW one with the same redshift obtained from the reconstructed result, and  then  all the available data points can be used to probe the cosmic opacity.  Furthermore, we discuss and obtain  the continuous redshift-dependent cosmic optical depth $\tau(z)$  by comparing the reconstructed LD from GW  with the reconstructed one from SNIa data.

The results show that the best value of  the continuous cosmic optical depth $\tau(z)$  varies from a small negative value to a small positive one in the redshift region $0.1<z<1.4$ as the redshift varies, although no deviation from a  transparent universe is found  at $1\sigma$ CL. We also obtain that the error bar of cosmic opacity  $\sigma_{\epsilon}\sim 0.011$ and $\sigma_{\epsilon}\sim 0.006$ for JLA and Pantheon respectively in a cosmological-independent way.  Compared with the previous actual tests, our result is nearly $80\%$  smaller than that obtained by using the angular distances  from various astronomic observations,  and it is nearly  $60\%$  smaller than that obtained by using LD  from future GW measurements in Refs.~\cite{JUN-JIE WEI2019,Jing-Zhao Qi2019} with a cosmological-independent  method. It can be concluded that, given the GW detectors are expected as the program of ET, future measurements of GW may offer  competitive constraints on the cosmic opacity. Therefore, future GW measurement may provide us with an opportunity to investigate  the spatial homogeneity of the cosmic transparency, and it can be considered as a powerful tool to probe cosmic opacity.

\begin{acknowledgments}
We very much appreciate helpful comments and suggestions from anonymous referees, and  helpful discussion from Mr. Puxun Wu.  We also like to thank for helpful discussion about the improvements of simulation process of GW data with Mr. Xuewen Liu and the improvements of  Gaussian Process in Python  with Mr. Zhengxiang Li.
This work was supported by the National Natural Science Foundation of China
under Grant Nos. 11147011, 11405052,11465011,and 11865011,  the
Hunan Provincial Natural Science Foundation of China under Grant No. 12JJA001,
the Foundation of Department of science and technology of Guizhou
Province of China under Grants No. J [2014] 2150 and the Foundation of
the Guizhou Provincial Education Department of China under Grants
No. KY [2016]104.

\end{acknowledgments}

\end{document}